\documentstyle[11pt]{article}
\newcommand{\asla}{\mbox{\ooalign{\hfil/\hfil\crcr$a$}}}

\newcommand{\eq}{\label}
\newcommand{\msv}{\mbox{\boldmath $\zeta_m$}}

\newcommand{\va}{\mbox{\boldmath $a$}}
\newcommand{\vp}{\mbox{\boldmath $p$}}

\newcommand{\CA}{\mbox{${\tilde{C}}_{A}$}}
\newcommand{\CT}{\mbox{${\tilde{C}}_{T}$}}

\newcommand{\epsi}{\mbox{$\varepsilon$}}

\newcommand{\espsm}{\mbox{$\varepsilon_{sp-sym}$}}

\newcommand{\tpsp}{\hspace{1.5em}}

\newcommand{\stret}[1]{\hbox{$\vcenter to #1{}$}}

\def\sigm{\mbox{$\langle \sigma \rangle$}}
\def\gs{\mbox{$g_\sigma$}}
\def\gv{\mbox{$g_\omega$}}
\def\ms{\mbox{$m_\sigma$}}
\def\mv{\mbox{$m_\omega$}}

\title{\bf  
Ferromagnetism of Nuclear Matter\\
in the Relativistic Approach }

\author{Tomoyuki~Maruyama$^{1,2,3}$ and Toshitaka~Tatsumi$^{4}$ \\
$^{1}$ {\it College of Bioresource Sciences,
Nihon University, Fujisawa, 252-8510, Japan} \\
$^{2}$ {\it RI Beam Science, RIKEN, Wako 351-0198, Japan} \\
$^{3}$ {\it Japan Atomic Energy Research Institute,
Tokai, Ibaraki 319-1195}\\
$^{4}$ {\it Department of Physics, Kyoto University
Kyoto 606-8502, Japan} }
\date{}

\begin{document}

\maketitle

\begin{abstract}
We study the spin-polarization mechanism in the highly dense nuclear matter
with the relativistic mean-field approach.
In the relativistic  Hartree-Fock framework we find that there are 
two kinds of spin-spin interaction channels, which are the axial-vector and 
tensor exchange ones.
If each interaction is strong and different sign, 
the system loses the spherical symmetry and holds
the spin-polarization in the high-density region.
When the axial-vector interaction is negative enough,
the system holds ferromagnetism.
\end{abstract}


\vfil
\eject

\newpage

\section{Introduction}

\tpsp
Resent discovery of  ''Magnetar'' \cite{magnetar}, which is a neutron star 
with super strong magnetic field, seems to revive 
a big question on the origin of the strong magnetic field. 
Since there is spread bulk hadronic
matter beyond the nuclear density inside neutron stars, it should be
interesting to consider its origin in the context of dynamics of
hadronic matter; e.g.,  
if the spin-polarization of baryons are realized in nuclear matter, 
ferromagnetism may occur in neutron stars. 

For quark matter, one of the authors (T.T.) has recently 
indicated 
a possibility of spin-polarization of quarks interacting with
one-gluon-exchange (OGE) interaction \cite{SP-QM}; 
the Fock exchange interaction between
quarks has a role to align spins, which is similar to the
electron system \cite{blo}. Using this result he suggested that
super strong magnetic field expected in magnetars may be explained if
they are quark stars.
There he also found {\it relativistic effects} give rise to a new mechanism 
for ferromagnetism, which is never appeared in the nonrelativistic case.

As for the normal nuclear matter Pandharipande et al. \cite{Pand}
have reported no possibility of stable spin-polarization 
within the non-relativistic
framework; magnetic susceptibility never changes its sign within
densities relevant for neutron stars. 

On the other hand Niembro et al. \cite{Niem1,Niem2} have suggested a 
possibility 
of spin-polarized nuclear matter using the relativistic Hartree-Fock 
(RHF) approach  \cite{RHF},
though spontaneous spin-polarization occurs at too high density. 
The results in Refs. \cite{Niem1,Niem2} suggest that the relativistic framework
may be  more favorite for spin-polarization than the nonrelativistic one.
Checking their framework we find some problems about the calculation.
First, they implicitly defined the spin-polarization of the system by
using the eigenstates of the spin operator 
$\Sigma_z (= \sigma_z \otimes 1) = \gamma_5\gamma_0\gamma^3$ in the
rest frame of each particle, and all the particles take the same
eigenvalues in the fully polarized state. 
This may be a direct analogue of the
nonrelativistic ferromagnetism. However, the spin operator cannot
commute with Hamiltonian in the relativistic theories and thereby we
must carefully treat two polarization degrees of freedom for baryons. 
Actually their choice is not a unique choice; the spin states of all
the baryons do not necessarily need to be 
the eigenstates of the spin operator
even in their rest frames. Instead, the spin orientation should depend on 
the momentum of each particle. Thus we must consider the spin
configuration 
of the spin-polarized state in the phase space.

Secondly, they kept spherical symmetry during the formulation
though the spin-polarized system may break it due to the existence of 
a specific direction (we can assign it as the $z$ axis without loss of
generality) along the nonzero magnetization vector in the
spin-polarized state. The interaction-energy density has generally
the additional momentum dependence besides that from the propagator 
in the relativistic theories due to the lower
component of the Dirac spinor.
Since there should be appeared a 
characteristic direction 
in the
spin-polarized state, there is another scalar product of momentum and
the unit vector along the $z$ axis besides that of momenta in the
interaction energy density; 
the expression of the interaction energy density is no longer
rotation-invariant and retains only the rotational symmetry around 
the $z$ axis. 
This may in turn suggest a possibility of axially symmetric deformation 
of the Fermi sphere. 

Thirdly, they used in Refs.\cite{Niem1,Niem2} 
the $\sigma$, $\omega$, $\rho$ and $\pi$-meson
exchange for the RHF calculation.
In reality these interactions should be considered to be the in-medium one, 
not the bare one,
and we do not have any enough information for the two-body interaction
which should be used there, especially for the spin-spin interaction
channels.

Putting these remarks aside, we know 
there has been no systematic and sufficient discussions on this topic 
in the relativistic many-body approach \cite{Serot},
particularly in view of the relativistic effects.
For example, the usual calculation, either in the non-relativistic
or relativistic framework,  has been done under the assumption of 
spherical symmetry for the mean fields.
In this paper we reexamine the spin-polarization of nucleon matter
within the RHF approach, focusing on the breaking down of spherical 
symmetry and importance of the relativistic effects. 
Since it has been shown in Refs.\cite{SP-QM,Niem1,Niem2} that the
Fock exchange interaction plays a key role in the context of the spin
polarization, we must treat this matter within the RHF approach. 
Then we take some choices of the spin-vectors 
dependent on the momentum of each particle, and study
relations between nuclear properties and these choices.

In the next section we give our formalism,
where we clarify what kinds of interactions are effective
for the spin-polarization and figure out the role of
the spherical symmetry breaking. 
We also classify the various spin configurations in the relativistic formulation, 
whereas there is a unique choice in the nonrelativistic theories.
Results  of numerical calculations are given in Sec. 3.
Sec. 4 is devoted to summary and concluding remarks.

\section{Formalism}

\tpsp
In this section we briefly explain our formulation to describe 
the spin-polarized system.
There should appear a special direction along the spin-polarization; 
it is defined to be oriented to the positive $z$-direction.
Such a system breaks spherical symmetry while the axial symmetry
around the $z$-axis is preserved.

In the RHF framework the interaction energy density 
in the isospin saturated system is generally written as
\begin{eqnarray}
\epsilon_{int} & = & 
\frac{1}{2} \int \frac{d^4p}{(2 \pi)^3} \frac{d^4k}{(2 \pi)^3} d^4k 
[~ Tr \{i S(p) \} {\cal D}_{S}(p-k) Tr \{i S(k)\} 
\nonumber \\ & &
+ Tr \{i S(p) \gamma_{\mu} \} {\cal D}_{V}(p-k) Tr \{i S(k) \gamma^{\mu}\}
\nonumber \\ & &
+ Tr \{i S(p) \gamma_{5}\} {\cal D}_{P} (p-k) Tr \{i S(k) \gamma_{5} \}
\nonumber \\ & &
+ Tr\{i S(p) \gamma_{5} \gamma_{\mu}\} {\cal D}_{A}(p-k) 
Tr\{i S(k) \gamma_{5} \gamma^{\mu}\}
\nonumber \\ & &
+ Tr\{i S(p) \sigma_{\mu \nu}\} {\cal D}_{T}(p-k) 
Tr\{i S(k) \sigma^{\mu \nu}\} ~]
\label{edint}
\end{eqnarray}
for the one-boson exchange type interaction, assuming 
no derivative coupling.  
Here $S(p)$ is the nucleon propagator with momentum $p$,
${\cal D}_\alpha$ $(\alpha = S, V, P, A, T)$
is the linear combination of meson-propagators
with the nucleon-meson couplings.
Note that these interaction terms do not necessarily appear in the
original Lagrangian,
and that some of them  are given by other ones by way of 
 the Fierz transformation.
When using the $\sigma$- (scalar), $\omega$- (vector) 
and $\eta$- (pseudo-scalar) meson exchanges, for example,
\begin{eqnarray}
{\cal D}_{S} (q) & = & - \frac{g_{\sigma}^2}{m_{\sigma}^2}
+ \frac{1}{8} g_{\sigma}^2 \Delta_{\sigma}(q) 
- \frac{1}{2} g_{\omega}^2 \Delta_{\omega}(q)
+ \frac{1}{8} g_{\eta}^2 \Delta_{\eta}(q) 
\label{DS}
\\
{\cal D}_{V} (q) & = & \frac{g_{\omega}^2}{m_{\omega}^2}
+ \frac{1}{8} g_{\sigma}^2 \Delta_{\sigma}(q) 
+ \frac{1}{4} g_{\omega}^2 \Delta_{\omega}(q)
- \frac{1}{8} g_{\eta}^2 \Delta_{\eta}(q) ,
\label{DV}
\\
{\cal D}_{P} (q) & = & 
- \frac{g_{\eta}^2}{m_{\eta}^2}
+ \frac{1}{8} g_{\sigma}^2 \Delta_{\sigma}(q) 
+ \frac{1}{2} g_{\omega}^2 \Delta_{\omega}(q)
+ \frac{1}{8} g_{\eta}^2 \Delta_{\eta}(q) ,
\label{DPS} \\
{\cal D}_{A} (q) & = & 
{\frac{1}{8}} g_{\sigma}^2 \Delta_{\sigma}(q) 
- \frac{1}{4} g_{\omega}^2 \Delta_{\omega}(q)
+ \frac{1}{8} g_{\eta}^2 \Delta_{\eta}(q) ,
\label{DAVHF}
\\
{\cal D}_{T} (q) & = & 
{\frac{1}{8}} g_{\sigma}^2 \Delta_{\sigma}(q) 
- \frac{1}{8} g_{\eta}^2 \Delta_{\eta}(q) ,
\label{DTHF}
\end{eqnarray}
where $g_a$ (a={$\sigma$}, {$\omega$} and {$\eta$}) is
the nucleon-meson coupling strength, and $\Delta_a$ is the meson-propagator,
\begin{equation}
\Delta_{a} (q) = \frac{1}{m_{a}^2 - q^2} ~~~~ (a=\sigma, \omega) .
\end{equation}
The first constant term {$g_\sigma^2/m_\sigma^2$} 
({$g_\omega^2/m_\omega^2$})
in {${\cal D}_S$} ({${\cal D}_V$}) indicates 
the Hartree direct contribution,
and other terms are the Fock exchange contributions.
Introducing other mesons or vertex form-factors to couplings
do not change the form of eq. (\ref{edint}).

In the spin-polarized system, the expectation value of the spin
operator $\Sigma_z$ has a nonvanishing value,
and thereby  
at least the axial-vector (A) and tensor (T)
exchange terms survive in eq. (\ref{edint}) as well as 
the scalar (S) and vector (T) terms. 
Then the nucleon self-energy must include their contributions,
and the ground state breaks parity and spherical symmetries.

In the relativistic mean-field (RMF) approach we usually neglect 
the momentum dependence of the propagator  
because the nucleon-nucleon interaction can be effectively treated as
the zero-range one in low energy phenomena as far as the typical
energy and/or momentum scale is much less than the meson mass.
Actually only very small momentum dependence has appeared in 
the full RHF calculation \cite{RHF,soutome}.
Here, we take the zero-range approximation for two nucleon
interaction which can be described as follows:
\begin{equation}
{\cal D}_{\alpha} =  \frac{{\tilde C}_\alpha}{2 M^2} ~~~~ 
(\alpha = S, V, P, A, T) .
\end{equation}

Even with this approximation the RHF calculation is still complicated
because there appear the axial-vector and tensor mean fields.
These parity violation terms largely mix the positive-energy 
and negative-energy states in the single particle wave-function.
Thus completely self-consistent calculations become very complicated
and must be done very carefully as for the vacuum polarization.
Furthermore we do not have sufficient information of
all channels of the  in-medium interaction between nucleons,
especially in the axial-vector and tensor channels.
Thus we will not get any clear conclusion in spite of very tough  calculations
by solving the RHF equation self-consistently.
Instead of solving the exact self-consistent RHF equation,
we take a variational approach in the RHF framework.

The two degrees of freedom of the spin polarization for each nucleon
is denoted by $\zeta=1$ and $\zeta=-1$, which we call spin-up and
-down, respectively.  
Then we take the nucleon propagator with four-momentum $p$ in the following 
form;
\begin{equation} 
S(p,\zeta) = S_F (p,\zeta) +  S_D (p,\zeta) 
\label{prop1}
\end{equation} 
with the propagators of a vacuum piece ($S_F$) and a density-dependent 
piece ($S_D$),
\begin{eqnarray} 
S_F(p,\zeta) & = & \frac{ \{ \gamma_{\mu} p^{*\mu} + M^*) \} }
                        {p^{*2} - M^{*2}} 
\frac{ \{ 1 + \gamma_5 \asla(p^*,\zeta) \} }{2} ,
\\
S_D(p,\zeta) & = & \{ \gamma_{\mu} p^{*\mu} + M^*) \}
\frac{ \{ 1 + \gamma_5 \asla(p^*,\zeta) \} }{2} 
\frac{i \pi}{E_p^*} n(\vp,\zeta) \delta(p_0 - \epsi_p),
\label{prop}
\end{eqnarray}
where $p^{*\mu}\equiv p^\mu-U^\mu$, $M^*=M-U_s$ and 
$E_p^* = \sqrt{\vp^2 + M^{*2}}$. 
In these equations $U_s$ and $U_\mu$ are the scalar 
and vector mean fields,
and $\epsi_p$ is the single particle energy of the nucleon with momentum 
{\vp}, $\epsi_p = E_p^* + U_0 $. 
$a(p^*,\zeta)$ is the spin-vector of the nucleon with momentum $p$ 
which satisfies the following conditions:
\begin{equation}
a_{\mu} a^{\mu} = -1,   ~~~~~  a_{\mu} p^{*\mu} = 0.
\end{equation}
In the following we only keep the density-dependent piece $S_D$ with the
momentum distribution function $n(\vp,\zeta)$ to be determined, 
for which we assume the axial symmetry along the spin-polarization. 
Note that this form of the propagator is the same as the one when 
the nucleon self-energy includes only scalar 
{$U_s$} and vector {$U_{\mu}$ mean fields,
which  are independent of the nucleon's spin. 
Moreover, we can easily
see that the expectation value of the pseudoscalar operator
automatically vanishes,
\begin{equation}
Tr\{ iS(p)\gamma_5\}=0.
\end{equation}
So the pseudoscalar (P) term in Eq.~(\ref{edint}) vanishes in this case.

Then the total energy density $\epsilon_{T}$
is separated into two parts: the spin-independent part  $\epsilon_{SID}$
and the spin-dependent part  $\epsilon_{SD}$ as
\begin{equation}
\epsilon_{T} = \epsilon_{SID} + \epsilon_{SD}.
\end{equation}
The axial-vector and tensor exchange channels contribute
to the spin-dependent part $\epsilon_{SD}$ while
the kinetic energy and contributions from the scalar and vector channels 
are involved in the spin-independent one   $\epsilon_{SID}$.

Under the zero-range approximation
the spin-independent part of the Fock contribution
can be incorporated into the Hartree one.
Then it is possible to  
redefine the two-body scalar interaction by taking into account the 
Fock terms,
which corresponds to the usual relativistic Hartree (RH) approximation.
Thus we can write the scalar mean-field by the expectation value of
the $\sigma$ field {\sigm} as 
\begin{equation}
U_s = \gs \sigm ,
\end{equation}
and calculate the effective mass {$M^*$} with the usual RH approximation.
The expectation value $\sigm$ is given by the equation,
\begin{equation}
\frac{\partial}{\partial \sigm} {\widetilde U} [\sigm]
= \gs \rho_s = \gs \sum_{\zeta} \rho_s(\zeta), 
\label{RHeq}
\end{equation}
where the scalar density is defined as
\begin{equation}
\rho_s(\zeta) =  
\int \frac{d^3 p}{(2 \pi)^3} n(\vp;\zeta) \frac{M^{*}}{E_p^*},
\end{equation}
and ${\widetilde U} [\sigma]$ is the self-energy potential 
of the sigma-field, whose expression is given in Ref. \cite{TOMO1,K-con}.
\begin{equation}
\widetilde{U} [\sigma] 
= \frac{ \frac{1}{2} m_{\sigma}^2 \sigma^2 
+ \frac{1}{3} B_\sigma \sigma^3
+ \frac{1}{4} C_\sigma \sigma^4 } 
{ 1 + \frac{1}{2} A_\sigma \sigma^2 } \  .
\eq{sigself}
\end{equation}

In the RH approximation, furthermore, the spatial components of the vector 
mean-field are vanished, and the time component is given as
\begin{equation}
U_0 = \gv \rho_B = \gv \sum_\zeta \rho_B (\zeta),
\end{equation}
where $\rho_B (\zeta)$ is the baryon density contributed from 
nucleon with the spin suffix $\zeta$ as
\begin{equation}
\rho_B(\zeta) =  
2 \int \frac{d^3 p}{(2 \pi)^3} n(\vp;\zeta) .
\end{equation}

Under this formulation the spin-independent part of the energy density
{$\epsilon_{SID}$} becomes
\begin{equation}
\epsilon_{SID} = 2 \sum_{\zeta}
\int \frac{d^3 p}{(2 \pi)^3} n(\vp;\zeta) {E_p^*}
+ \widetilde{U} [\sigma] 
+ \frac{g_{\omega}^2}{2 m_{\omega}^2} [ \sum_{\zeta} \rho_B (\zeta) ]^2,
\end{equation}
while the spin-dependent energy density $\epsilon_{SD}$ is calculated   
to be,
\begin{eqnarray}
\epsilon_{SD} & = & 
\frac{\CA}{2 M^2} \rho^2_{A} +  \frac{\CT}{2 M^2} \rho^2_{T}
\label{engSD}
\end{eqnarray}
with the axial-vector and tensor densities,
\begin{eqnarray}
\rho_{A} & = & \int \frac{d^4 p}{(2 \pi)^4} 
Tr \{S_D(p) \gamma_5 \gamma^3 \} = \rho_B <\Sigma_z> = 
\sum_{\zeta} \int \frac{d^3 p}{(2 \pi)^3} n(\vp;\zeta) \frac{M^*}{E^*_p} a_z
\label{rho-AV} \\
\rho_{T} & = & \int \frac{d^4 p}{(2 \pi)^4} 
Tr \{S_D(p) \sigma_{1 2} \} = \rho_B <\beta \Sigma_z> = 
\sum_{\zeta} \int \frac{d^3 p}{(2 \pi)^3} n(\vp;\zeta) 
\{ a_{z} - \frac{p_z}{E^{*}_p} a_0 \}
\label{rho-T}
\end{eqnarray}
Other components of the axial-vector and tensor densities 
are vanished because of the axial symmetry of the momentum distribution.
Note that the axial-vector and the tensor  interactions, 
even though they are not necessarily 
included in the original  Lagrangian, may arise from the Fock
exchange interactions by way of the Fierz transformation, as is seen
in Eqs.~(5) and ~(6). In this sense we can say that the Fock exchange
interaction is essential for the system to be ferromagnetic 
\cite{SP-QM,Niem1,Niem2}.

In order to figure out the properties of the spin-polarized matter, 
we solve RH equation (\ref{RHeq})
and calculate the energy-density by fixing the baryon density
$\rho_B$ and the spin-polarization parameter $x_s$ defined by
\begin{equation}
x_s \equiv ( \rho_{\uparrow} -  \rho_{\downarrow} )/ \rho_B .
\end{equation}
where {$\rho_{\uparrow} = \rho_B (\zeta=1)$} and
 {$\rho_{\downarrow} = \rho_B (\zeta=-1)$}.
(For convenience the spin-up and spin-down states are indicated by 
the symbols 
$\uparrow$ and $\downarrow$, respectively.)

In our approach the wave-function is not an exact solution of  
the Dirac equation in the mean-fields.
So we need to specify the spin configuration in the system by choosing 
a spin-vector $a_\mu$. Once it is fixed, the momentum distribution
of the single-particle state should be 
also determined accordingly. 
The best way to this end is to choose the configuration 
to optimize the total energy density $\epsilon_{T}$, while it may be 
rather complicated. Instead, we examine here the following three choices by 
physical considerations.

The total spin-polarization is directed to the positive 
direction of the $z$-axis;
here we define a unit vector $\msv = (0,0,\zeta)$.
Usually the spin-vector $a_{\mu}$ is chosen as (0,{\msv}) at 
the rest frame of the nucleon, we should call this choice as Choice-1(Ch1).
This choice may be a natural extension from the nonrelativistic
ferromagnetism and has been also taken in the context of 
ferromagnetism of quark
matter \cite{SP-QM}.
Then the spin-vector with momentum {\vp} becomes
\begin{equation}
{\va} = [ {\msv} + \frac{(\msv \cdot \vp) \vp}{M^* (E_p^* + M^*)} ]
~,~~ a^0 = \frac{\msv \cdot \vp}{M^*}, 
\end{equation}
by way of the Lorentz transformation.
In this choice $\rho_{A}$ and $\rho_T$  can be written as
\begin{eqnarray}
\rho_{A} & = & \frac{1}{3} \sum_{\zeta} \zeta 
\{ 2\rho_s({\zeta}) + \rho_B({\zeta}) \} ,
\label{rhoAV-s}
\\
\rho_{T} & = & \frac{1}{3} \sum_{\zeta} \zeta 
\{ \rho_s({\zeta}) + 2 \rho_B({\zeta}) \} .
\label{rhoT-s}
\end{eqnarray}

Then the interaction-energy density can be written only in terms of 
the scalar {$\rho_{s}(\zeta)$} and vector densities {$\rho_{B}(\zeta)$}
of nucleons with the spin {$\zeta$}, 
and the expression still holds spherical symmetry.
In the high-density limit, as mentioned before, 
the scalar density $\rho_s$ approaches to a finite value 
in the RMF approach \cite{Serot}, 
and its contribution becomes negligible, so that we can see that $
\rho_{A}^2 < \rho_{T}^2$ in the limit $\rho_B \rightarrow \infty$.
Hence the spin-polarization occurs when ${\CA}+4{\CT}<0.$
The above choice does not necessarily lead to the state with the
spin-alignment, and
the total spin per nucleon 
converges to one-third 
($< \Sigma_z /A > = \rho_{A}/\rho_B \rightarrow x_s/3$)
in the infinite density limit; magnetization is not so large
even if all the nucleon spins align.

Here we consider another choice, Choice-2(Ch2) for the vector $a_\mu$ to 
give the maximum for $|a_z|$ within our framework.
In this new choice the spin-vector becomes
\begin{eqnarray}
a_0 & = & \frac{E_p^* (\msv \cdot \vp)}{M^* \sqrt{(\msv \cdot \vp)^2 + M^{*2}}} , \\
{\va} & = & \frac{ M^{*2} \msv + (\msv \cdot \vp) \vp }
{M^* \sqrt{(\msv \cdot \vp)^2 + M^{*2}}} 
\label{s-dir}
\end{eqnarray}
Substituting the above form into  
eqs.(\ref{rho-AV}) and (\ref{rho-T}), 
we get 
\begin{eqnarray}
\rho_{A} & = & 2 \sum_{\zeta} \zeta \int \frac{d^3 p}{(2 \pi)^3}
n(\vp;\zeta) \frac{\sqrt{p_z^2 + M^{*2}}}{E_p^*} 
\label{rhoAV}
\\
\rho_{T} & = & 2 \sum_{\zeta} \zeta \int \frac{d^3 p}{(2 \pi)^3}
n(\vp;\zeta) \frac{M^{*}}{\sqrt{p_z^2 + M^{*2}}} 
\label{rhoTN}
\end{eqnarray}
When $\rho_B \rightarrow \infty$,
$\rho_{T}$ converges to a finite value,
and  $\rho_{A}$ is proportional to the baryon density $\rho_B$.
If the momentum-distribution $n(\vp;\zeta)$
is taken as usual Fermi-distribution,
the total spin per nucleon converges 
\begin{equation}
< \Sigma_z /A > = \rho_{A}/\rho_B {\rightarrow} x_s/2.
\end{equation}
Hence we can expect a ferromagnetic phase as long as ${\CA}<0$.
We shall see 
that this choice is the most appropriate for ferromagnetism.

As another choice,  Choice-3(Ch3), we take the vector $a_\mu$ 
to give the maximum 
of the expectation value $< \beta \Sigma_z /A > = \rho_{T}/\rho_B$, 
which is reduced to the same value as that of the spin
operator in the nonrelativistic limit.
In this case the spin-vector becomes
\begin{eqnarray}
a_0 & = & \frac{E_p^*}{M^* \sqrt{\vp_T^2 + M^{*2}}} \zeta, \\
\va & = & \frac{\sqrt{\vp_T^2 + M^{*2}}}{M^*} \msv, 
\label{sB-dir}
\end{eqnarray}
with
\begin{equation}
{\vp_T} = \vp - \msv ( \msv \cdot \vp ).
\end{equation}
Then the axial-vector and tensor densities can be written as 
\begin{eqnarray}
\rho_{A} & = & 2 \sum_{\zeta} \zeta \int \frac{d^3 p}{(2 \pi)^3}
n(\vp;\zeta) \frac{M^{*}}{\sqrt{p_T^2 + M^{*2}}} \\
\rho_{T} & = & 2 \sum_a \zeta \int \frac{d^3 p}{(2 \pi)^3}
n(\vp;\zeta) \frac{\sqrt{p_T^2 + M^{*2}}}{E_p^*}.
\end{eqnarray}
In this choice 
$\rho_{A}$ and $\rho_{T}$ show the opposite behaviors  
in the infinite density limit;
$\rho_{A}$ converges to a finite value,
and  $\rho_{T}$ is proportional to the baryon density $\rho_B$. Hence
the system becomes spin-polarized state when ${\CT}<0$. 
However, the expectation value of the spin operator $\Sigma_z$
has a nonvanishing value
and becomes
\begin{equation}
< \Sigma_z /A > = \rho_{A}/\rho_B {\rightarrow} 0 ,
\end{equation}
which means that the system is not ferromagnetic in a usual sense.

Since the spin-dependent energy $\epsilon_{SD}$ 
depends on the coupling strengths, {\CA} and {\CT}, we have to
carefully determine  which choice is most appropriate.
In the low density region  $\rho_{A} \approx \rho_{T} \approx x_s \rho_B$,
the spin-saturated system should be stable and then the two coupling constants
\CA~ and \CT~ must satisfy the following relation
\begin{equation}
\CA + \CT > 0 .
\end{equation}
If {$\CA \ge 0$} and {$\CT \ge 0$},
the spin-saturated system must be stable in all the density region, 
and spherical symmetry is always held. 
Otherwise the analysis at the infinite density limit show us that
system  becomes spin-polarized
if $\CA < 0 < \CT$ (Ch2) or $\CT < 0 < \CA$ (Ch3) holds.
Even in Ch1 the spin-polarization can occur if $\CA + 4\CT <0$, 
but this condition expects large negative {\CT}  and the spin-vector of a nucleon prefers Ch3 than Ch1.

Furthermore we should note that the above expressions of 
eq.(\ref{rhoAV}) and eq.(\ref{rhoTN}) do not preserve spherical symmetry.
The relativistic effects automatically give rise to the spherical symmetry
breaking.
From this fact we can naturally expect that the momentum distribution
$n(\vp;\zeta)$ is allowed to be distorted while keeping the axial-symmetry.
In order to estimate the effects of distortion of the momentum distribution,
we introduce the quadrupole-distorted distribution function $n(\vp)$ as
\begin{equation}
n(\vp; \zeta) =
n_{0}( e^{\lambda(\zeta)}p_x,~e^{\lambda(\zeta)}p_y,~ 
e^{-2\lambda(\zeta)}p_z;~\zeta ),
\end{equation}
where $n_0(\vp; \zeta) = \theta(p_F(\zeta) - |\vp|)$  with the Fermi-momentum
{$p_F$}.
The parameter $\lambda(\zeta)$ is determined to give the energy minimum
of the spin-polarized system. 

As mentioned in the previous section, the two contributions from 
the axial-vector and
tensor channels are not considered in the original interactions; 
instead they are derived by the Fierz transformation from the Fock 
exchange interactions in other channels.
Using several parameters given in previous works, however,
we get  various spin-properties even at the normal nuclear density. 
For example the parameter-sets of  HF-I in Ref. \cite{RHF},
within the {$\sigma$-} and {$\omega$-} meson exchanges,
gives us $\CA = -8.55$ and $\CT = 30.2$,
and the parameter-sets with $\pi$-, $\sigma$-, $\rho$- and $\pi$(PS)- meson exchanges
in Ref. \cite{Niem2},
gives us $\CA = 3210$ and $\CT = -3200$.
The latter extraordinary value comes from the pion-exchange
($M^2 g^2_{\pi} / m^2_{\pi} = 8200$), which
never allow the zero-range interaction approximation.
Thus the values of two coupling strengths, {\CA} and {\CT}, are still 
very ambiguous, and 
they cannot be individually determined at present 
both in theoretical and experimental ways.
Instead of using parameters given in previous papers, 
we investigate the spin-polarization of nuclear matter by 
varying the values of \CA~ and \CT ~in this paper.

As discussed above, we have a possibility to get ferromagnetic matter
only in the case of {$\CA < 0 < \CT$}, where Ch2 for the spin-vector
must be most appropriate.
Hence we study the spin-properties only in this case except in some
figures where properties are also calculated with Ch1 for comparison.
We shall see that ferromagnetism can occur due to the spherical 
symmetry breaking in the relativistic framework, 
by concrete numerical calculations in the next section.

\section{Results and Discussions}
 
\tpsp
In this section we make concrete calculations as for 
the spin-polarized nuclear matter and discuss their consequences. 
We calculate physical quantities concerning the magnetic properties 
only for the isospin symmetric matter and make comparison with neutron 
matter in the final place.
In the actual procedure, first, we evaluate 
the sigma-field  with eq.(\ref{RHeq}) by fixing baryon density $\rho_B$ 
and the spin-polarization parameter 
$x_s \equiv ( \rho_{\uparrow} -  \rho_{\downarrow} )/ \rho_B$;
namely we solve the RH equation.
Secondly, we substitute the result into eqs. (\ref{rhoAV}) 
and (\ref{rhoTN}), 
and obtain the spin-dependent energy $\epsilon_{SD}$ in eq.(\ref{engSD}).
Repeating these processes by varying 
$\lambda_{\uparrow} \equiv \lambda (\zeta=1)$ and 
$\lambda_{\downarrow} \equiv \lambda (\zeta=-1)$, 
we search the energy minimum.

The parameter-sets PM1 ($M^*/M = 0.7$) and PM4  ($M^*/M = 0.55$) 
are used for the RH calculation;
the definite values of the parameters are given in Table 1.
In Fig. 1 we show the density-dependence of the total energy per nucleon 
($E_{T}/A$) and the effective mass normalized by the bare mass ($M^*/M$).
The values of the coupling strengths in the axial-vector and tensor
channels  must be consistent with the spin-properties
at normal nuclear density such as the spin-symmetry energy.
We here define the spin-symmetry energy, inversely proportional to the 
magnetic susceptibility by
\begin{equation}
\espsm =
 \frac{\partial^2 E_{T}/A}{\partial <\Sigma_z/A>^2} |_{x_s = 0} .
\end{equation}
In this work we take its value as $\espsm = 25$(MeV),
while it is not clearly determined from experimental information
\footnote{The spin-symmetry energy is written as 
$
\espsm = ({p_F^2}/{6 m^*})(1 + G_0)
$
with the nonrelativistic effective mass $m^*$ and the Landau parameter 
$G_0$.
This value is given as 26 MeV by a realistic nuclear matter
calculation \cite{Muether} and 10 $-$ 60 MeV by the Skyrme
interaction \cite{Chabanat,TomoTsuka,Navarro}.},
and we use three kinds of the parameter-set:
$\CA = 0$ (SD1),  $\CA = -50$ (SD2) and $\CA = -100$ (SD3).
Here we restrict ourselves to the cases with $\CA < 0$ , since 
matter would be ferromagnetic only in this case.
The detailed values of parameters are given in Table 2.

In Fig. 2 we show the density-dependence of 
the spin-symmetry energies  {$\espsm$} with PM1 (a) and PM4 (b).
If this value becomes negative, the spin-saturated system becomes unstable
and the spin-polarized one is favored. 
The long-dashed, dashed and solid lines indicate results
for SD1, SD2 and SD3, respectively.
For all the parameter-sets the spin-symmetry energy increases 
 in the low density region as baryon density becomes larger.
While the spin-symmetry energy monotonously increases in the
case of $\CA=0$ (SD1),
it decreases and 
becomes minus above a critical density $\rho_c$
for the cases of negative {\CA} (SD2 and SD3):
$\rho_c/\rho_0 =8.74$ for SD2 and $\rho_c/\rho_0 =4.28$ for SD3.
Note that the spin-symmetry energies are smaller 
in high-density region in PM4 than in PM1,
The parameter-set PM4 gives rise to a smaller effective mass than PM1, 
and the decrease of the effective mass enlarges {\espsm} 
if the coupling strengths {\CA} and {\CT} are fixed.
Since we fix the spin-symmetry energy at the saturation density, however, 
the coupling constant {\CT} becomes small for the small effective mass;
then contributions from the axial-vector channel get larger in PM4.

In the infinite density limit the effective mass $M^*$ goes to
zero (Fig. 1b) and the scalar-density $\rho_s$ converge to the finite
value in the RMF theory.
Thus  $\rho_{A}$ and $\rho_{T}$ must have density-dependence similar
to  $\rho_{B}$ and $\rho_{S}$, respectively;
namely, when $\rho_B \rightarrow \infty$,
$\rho_{T}$ converges to a finite value
and  $\rho_{A}$ is proportional to the baryon density $\rho_B$.
In this limit, then, only the kinetic energy and the axial vector exchange
channels contribute to the spin-symmetry energy {\espsm}.
Since {$<{\Sigma_z}/A> \sim x_s$} around the spin-saturated matter, 
the contribution from the kinetic energy is proportional 
to {$\rho_B^{2/3}$} in the low density limit and 
to {$\rho_B^{1/3}$} in the high-density limit.
On the other hand the contribution from the axial-vector exchange channels
is proportional to {$\CA \rho_B$} in the high-density limit;
this behavior should be the same as 
the density-dependence of the (isospin-) symmetry energy. 
From this fact we can easily see that, if {$\CA<0$}, 
the spin-symmetry energy becomes negative
and the spin-polarization spontaneously occurs 
above a certain critical density.

In Fig. 3 some quantities are shown as 
functions of the spin-polarization parameter 
$x_s$
at $\rho_B = \rho_0$ (dotted line), $3\rho_0$ (dashed line), $5\rho_0$ 
(solid line) and $6\rho_0$ (chain-dotted line).
We give the energy difference from that 
at the spin-saturated matter ${\Delta E_T}/A = (E_T({x_s}) - E_T({x_s}=0))/A$
(a), $M^*/M$ (b),
the total spin per nucleon $< \Sigma_z /A > = \rho_{A}/\rho_B$ and 
the deformation parameters {$\lambda_{\uparrow}$} and {$\lambda_{\downarrow}$}.

In Fig. 3a it can be seen that above $\rho_B > 5 \rho_0$ the value of
the spin-polarization parameter at the energy-minimum moves 
from $x_s = 0$ to a finite value, whose
value becomes larger as baryon density increases.
There is a single local energy-minimum at the fixed density.
Thus the phase transition from normal matter to
the spin-polarized one is  of the second order.
\footnote{It should be interesting to compare this result with the
one in quark matter \cite{SP-QM}.}
To occur the first order phase-transition we need an interaction
energy which is negative and proportional to at least $< \Sigma_z /A >^4$.
The axial-vector term in the interaction energy  has the role to 
give rise to a spin-polarization, and it is exactly proportional to 
$< \Sigma_z /A >^2$.
On the contrary we can see in Fig. 3b that the effective mass is very
slightly varied with 
the spin-polarization parameter, and thereby we can conclude that the scalar 
interaction does not
largely contribute the spin-polarization.

In Fig. 3c  we can see that at high density the value of 
$<\Sigma_z /A> = \rho_A/\rho_B$ 
becomes slightly smaller than that in the fully polarized system . 
From eq. (\ref{rhoAV}) we can easily understand this behavior by
considering 
that the small effective mass reduces the expectation value of the nucleon spin
due to the relativistic effect ($M^*/E^*_p < 1$).

In Fig. 3d it can be seen that at high density the momentum distribution
gets the prolate deformation for spin-up nucleons  and the oblate deformation 
for spin-down nucleons; these deformations enhance $\rho_{A}$
(\ref{rhoAV}) and decrease $\rho_{T}$ (\ref{rhoTN}).

In Fig. 4 we show the equation of states for the fully spin-polarized 
nuclear matter ($x_s = 1$), 
the density-dependence of $\Delta E_{T} /A$ (a), $<\Sigma_z/A>$ (b) and
the deformation parameters for the spin-up nucleons $\lambda(\zeta=1)$ (c),
with the three spin-dependent parameter-sets (SD1$-$3).
As increasing {\CA} the deformation becomes larger, keeping rather
large value
of $<\Sigma_z /A>$, which in turn implies that 
the phase transition more easily occurs. 

When $\CA < 0$, the axial-vector exchange interaction plays a role to decrease
the total energy by enlarging $<\Sigma_z /A>$ with the variation of
$x_s$ and $\lambda_{\uparrow (\downarrow)}$.
On the other hand these variations enhances the kinetic
and scalar exchange energies;
the latter is not seen to be so effective because the effective
mass slightly changes from the spin-symmetric value (see Fig. 3b).

In order to examine effects of the spherical symmetry breaking,
in Fig. 5, we show the spin-symmetry energy {\espsm}
using three kinds of choices (Ch1, Ch2-S, Ch2-Q). Ch2-S and -Q are the 
two versions of Ch2: the spin-vector of Ch2 (eq.(\ref{s-dir})) 
with the spherical (Ch2-S)  
and the quadrupole-deformed momentum distribution (Ch2-Q).

In Ch1 the spin-symmetry energy always monotonously increases
when density becomes larger; the reason has been given in the previous
section.
In Ch2-S the qualitative behavior is almost the same as that for Ch2-Q
though the value of {\espsm} and the critical density are always
larger than those for Ch2-Q. 
When $\CA < 0$, the spherical symmetry breaking for the spin-vector
makes a critical effect for the spin-polarization.
In addition such effects from the choice of the spin-vector 
are enhanced  by the deformation of the momentum distribution (Fig. 4c). 
Hence the axial-vector correlation between two nucleons rather easily
gives rise to the ferromagnetic state through the spherical symmetry breaking.

In Fig. 6 we show the density dependence of 
{$< \Sigma_z /A > = \rho_{A}/\rho_B$} (a) 
and {$<\beta \Sigma_z /A> = \rho_{T}/\rho_B$} (b) at {$x_s = 1$} 
using Ch1 (dotted line), Ch2-S (dashed line) and  Ch2-Q (solid line),
and the quadrupole deformation parameter for spin-up nucleons
$\lambda_{\uparrow}$ using Ch2-Q (c). 
The parameter-sets PM1 and SD1 are used in this calculation.

With increase of density, the difference among three choices  
becomes prominent:
the total spin per nucleon {$<{\Sigma_z}/A>$} decreases for Ch1 and Ch2-S,
while it does not become so small for Ch2-Q (see Fig. 7a). 
This behavior can be understood from the analysis at the infinite
density limit.
In this limit, as mentioned before,
the effective mass approaches to the zero value,
and thereby  
$<{\Sigma_z}/A> \rightarrow {x_s}/3$ for Ch1 (see eq.(\ref{rhoAV-s}))
and
$<{\Sigma_z}/A> \rightarrow {x_s}/2$ for Ch2-S (see eq.(\ref{rhoAV}));
in any choice of the spin-vector the total spin, in the relativistic
framework, 
becomes much less than that in the non-relativistic one.
On the other hand the choice Ch2-Q gives 
\begin{equation}
<{\Sigma_z}/A> \rightarrow \frac{1+x_s}{2(1+e^{-6\lambda_{\uparrow}})}
- \frac{1-x_s}{2(1+e^{-6\lambda_{\downarrow}})}.
\end{equation}
In the limit of $\lambda_{\uparrow} \rightarrow \infty$ and
$\lambda_{\downarrow} \rightarrow -\infty$,
$<{\Sigma_z}/A> \rightarrow (1+x_s)/2$;
of course this limit makes infinite kinetic energy so that
the $\lambda_{\uparrow(\downarrow)}$ finally has a moderate value.
Thus the deformation of the momentum distribution plays a significant role
to give a large value of the total spin.
The prolate deformation of the momentum distribution 
recovers the reduction of the total spin in the high-density region. 

Furthermore it can be seen in Fig. 6b that
{$<{\beta \Sigma_z}/A>$} decreases for Ch2-S and Ch2-Q with increase of 
density, while its value does not becomes small.
As discussed in the previous section,
$<{\beta \Sigma_z}/A> \rightarrow 2{x_s}/3$ for Ch1 and
$<{\beta \Sigma_z}/A> \rightarrow 0$ for Ch2-S and Ch2-Q 
in the infinite density limit ($M^* \rightarrow 0$).
In addition we see that the value of $<{\beta \Sigma_z}/A>$ in Ch2-Q is always
smaller than that in Ch2-S, and we know here again that 
the deformation of the momentum distribution plays a important role 
through the reduction of $<{\beta \Sigma_z}/A>$.

From these results we confirm  that the choice of the spin-vector is 
very important for the spin-polarization, and
that we have to use Ch2 instead of Ch1 if $\CA < 0$.
In the Refs. \cite{Niem1,Niem2} the coupling {\CA} becomes positive,
though the interaction is not zero-range, and then Ch2 may
not be appropriate.
Then there remain Ch1 and Ch3; they used Ch1.

As made before, the analysis at the infinite density limit
must be useful to examine it qualitatively.
In this limit the spin-dependent part of the energy density becomes
\begin{eqnarray}
\epsilon_{SD} & \approx & \frac{1}{18} (\CA + 4\CT) \rho_B^2 x_s^2
~~~~~~~(in Ch1)
\\
\epsilon_{SD} & \approx & \frac{\pi^2}{32} {\CT} \rho_B^2 x_s^2
~~~~~~~~(in Ch3)
\end{eqnarray}
where we assume that the momentum-distribution holds the spherical symmetry.
To get smaller energy in Ch1 than in Ch3, 
\begin{equation}
\frac{1}{9}(\CA + 4\CT) < \frac{\pi^2}{16} {\CT}
\end{equation}
and this equation leads to a condition that $\CA < 1.55\CT$, which
is inconsistent with the condition $\CT < 0 < \CA$.
In addition, the deformation of the momentum distribution further 
reduces $\epsilon_{SD}$ in Ch3.

If $\CT < 0 < \CA$, then, Ch3 must become reasonable.
We can suppose that the 
qualitative behavior must be similar to that of Ch2-Q.
As the density becomes larger, namely, 
{$\rho_A$} converges to a finite value, and  
{$\rho_T$} increases proportionally to the density, and then the 
phase-transition occurs at a certain critical density.
In this case the momentum distribution is deformed with the oblate shape
for the spin-up ($\lambda_{\uparrow} < 0$) and 
with the prolate for the spin-down ($\lambda_{\downarrow} > 0$).
From this consideration we can expect that the total energy should be
smaller in Ch3 than that in Ch1 at any density if $\CT < 0 < \CA$.

From the above results we can see the particular role of 
the spherical symmetry breaking
through the choice of the spin-vector and 
the distortion of the momentum distribution. 
They becomes very important 
in high-density region due to decreasing of the nucleon effective mass. 

As mentioned in Sec. I the spontaneous spin-polarization is expected
to occur in the high-density region inside neutron stars,
though we have studied the magnetic properties only at the 
symmetric nuclear matter.
We did because we would like to clarify the discussion by avoiding 
extending the formulation 
to include the isovector channel.
We expect to determine the spin-spin interactions in the relativistic
framework from some experimental information in future.
However we may have some meaning to compare results at neutron matter
with that at symmetric nuclear matter within the formulation given 
in this paper.
In Fig. 7 we show the density dependence of the spin-symmetry energy
with PM1 and SD(1-3) for Ch2.
The thick and thin lines indicate results at neutron matter and
at symmetric nuclear matter, respectively.
We cannot see any significant difference between them
except that {\espsm} at neutron matter is a little larger than that at 
symmetric nuclear matter.

\section{Summary and Concluding remarks}

\tpsp
In this paper we have examined a possible mechanism of the 
spin-polarization of nucleons and discussed magnetic properties of the 
system.
In the relativistic framework there are two kinds of spin-spin interaction
channels, the axial-vector and tensor ones, which are reduced to the 
 same interaction
channels in the nonrelativistic framework. 
If the interaction energies from two channels have opposite signs, 
there is a second-order 
phase transition to a spin-polarized state.
Though the effects from  two channels are counterbalanced with each other 
around the normal density, the channel with the negative sign
becomes dominant, suppresses the spin-symmetry energy with increase
of density and induces a phase-transition in a certain critical density 
$\rho_c$.
In this mechanism the spherical symmetry breaking through the spin vector
and the momentum distribution plays a significant role;
the spherical symmetric calculation (Ch1) cannot describe
such a phase-transition  sufficiently.

These qualitative findings can be easily derived from the analysis at the
infinite density limit, which is equivalent to the ultra relativistic limit
in the present framework.
Actual numerical calculations confirm them,
so that the consideration in this limit is very useful 
to predict qualitative behaviors of matter at high density.

If the tensor channel is largely negative,  
the total nucleon spin becomes large while the total spin converges to zero 
at the infinite density limit.
Thus such system cannot make strong ferromagnetism even at high  density,
Inside actual neutron stars the density is not infinite, and
we has not known how large ferromagnetism is necessary to explain
magnetars.
Then we cannot deny this possibility though we omit it 
at the present calculations.
We should keep it in mind for future.

In this work we represent the strength of these channels with 
the couplings {\CA} and {\CT} by using the zero-range approximation,
which makes the Fock exchange interaction local.
Then we find that, if $\CA < -100$, the phase-transition occurs
in the reasonable density $\rho_B \le 5\rho_0$, which can be realized
inside neutron stars. However, 
the main purpose of this paper is to reveal the characteristic features 
of the relativistic ferromagnetism of nuclear matter within the RHF framework.
If we want to get a realistic  conclusion about the critical density, 
we need to determine two couplings {\CA} and {\CT} individually
from the experimental information, while 
experiments can at present give only nuclear spin properties in low density
region around and/or below the normal density. Furthermore, we need
the isovector interactions when we consider neutron-star matter.

Two approximations have been introduced in this work.
One is that we have applied a variational approach,  
avoiding the complete self-consistent calculation.
If we can do it, some points are improved; the single particle
energies become different between the spin-up and -down
states, which should in turn determine the modification of the Fermi sea.
Because of the variational principle, anyway, these improvements must reduce 
the total energy in the spin-polarized system.
Then the phase transition may occur in the lower density and/or
with the smaller couplings of {\CA} than those in the present calculations.

The other is that we have used the zero-range approximation for the meson
propagators and discarded the finite-range effects in the  
nucleon interactions;
the zero-range approximation is equivalent to neglect the 
momentum-dependence of the self-energies and should be reasonable, at
least, at low densities s.t. $m_a \gg p_F$.
As for the spin-independent parts 
it has been reported that the momentum-dependence of
the  self-energy has a role to reduce the total energy per nucleon 
($E_T/A$) in the high-density region \cite{TOMO1,KLW1}
and to suppress largely the Fermi velocity \cite{FV}, particularly 
in the low density region.
However these effects does not affect the nuclear equation of state 
at zero temperature.
As for the spin-dependent parts the situation must be similar.

These approximations used in our calculations must never spoil our qualitative
findings of the relativistic ferromagnetism.

We do not clearly know what phase actually appears inside neutron stars. 
A quark matter \cite{SP-QM} is one possibility and 
the hyperonic matter is also possible.
In the latter case we need to take into account the interaction
between hyperon and nucleon  and that between hyperons.

We compare here our results with the previous one given by one of the
authors (T.T.) for quark matter \cite{SP-QM}. It has been shown that 
ferromagnetism of quark matter may occur at low-density region and it
should be the first-order phase transition; all the quark spins 
suddenly align at the critical density and there is no partially 
polarized state. The most important difference from nuclear matter is
that quarks interact with each other through only the vector interaction 
by gluons and there is no direct interaction due to color neutrality of quark
matter. Moreover  there is no tensor channel 
and the gluon propagator
corresponds to the zero-mass limit of {$\Delta_{\omega}$}. As a result 
${\cal D}_A(q)$ given in Eq.~(5) is always negative and ${\cal
D}_T(q)\equiv 0$, which is most favorite situation for the system to be
ferromagnetic. In this case the momentum 
dependence of the propagator becomes essential and the phase transition is 
of the first order due to this effect; if we replace the propagator by 
some constant, we shall see that phase transition becomes of the
second order and there is a partially polarized state as is seen in
this paper. 

Our findings of the momentum dependence of the spin
orientation and possible deformation of the Fermi sea have not been 
taken into account in ref.\cite{SP-QM}. 
It may be interesting to see
these effects in the context of ferromagnetism of quark liquid \cite{ushi}.

\vskip 0.5cm
\centerline{\bf ACKNOWLEDGEMENTS}
\vskip 0.5cm
This work was supported in part by the Japanese Grant-in-Aid for
Scientific Research Fund of the Ministry of Education, Science, Sports 
and Culture (11640272).


%
\newpage

{\Large Figure Captions}

\bigskip

\begin{itemize}

\item[Fig. 1]
Density-dependence of the total energy per nucleon (a),
and the ratio of the effective masses to the bare masses for nucleon (b).
The solid and dashed lines indicate the results for PM1 and PM4,
respectively.

\item[Fig. 2]
Density-dependence of the spin-symmetry energies  ($\espsm$) with PM1 (a) 
and PM4 (b).
The long-dashed, dashed and solid lines indicate results
for $\CA=0$ (SD1), $=-50$ (SD2) and $=-100$ (SD3), respectively.

\item[Fig. 3]
Energy difference between the spin-polarized and saturated systems (a),
the effective mass $M^*$ normalize by the bare nucleon mass $M$ (b)
the total spin per nucleon {$< \Sigma_z /A >$} (c).
and the deformation parameter $\lambda$ (d) versus the
spin-polarization 
parameter {$(\rho_{\uparrow} - \rho_{\downarrow})/\rho_B$}. 
The dotted, dashed, solid and chain-dotted line indicate
results at {$\rho_B = \rho_0$},  {$\rho_B = 3\rho_0$},
{$\rho_B = 5\rho_0$}  and  {$\rho_B = 6\rho_0$}, respectively  
In the third column (c) the thick and thin lines indicate ones 
for the spin-up and spin-down, respectively.

\item[Fig. 4]
The density-dependence of $\Delta E_{T} /A$ (a), $<\Sigma_z /A>$ (b) 
and the deformation parameters for the spin-up nucleons 
$\lambda(\zeta=1)$ (c) with SD1 (dotted line),  
SD2 (dashed line) and  SD3  (solid line).
in the fully spin-polarized nuclear matter ($x_s=1$).

\item[Fig. 5]
Density-dependence of the spin-symmetry energies  ($\espsm$) with PM1
calculated with the method Ch1 (a), Ch2-S (b) and Ch2-Q (c) (seeing text). 
The meaning of the lines are shown in Fig. 2.

\item[Fig. 6]
The density-dependence of  $<\sigma_z /A>$ (a) $<\beta \sigma_z /A>$ (b)
and the deformation parameters for the spin-up nucleons 
$\lambda(\zeta=1)$  with SD1 (dotted line),  
SD2 (dashed line) and  SD3  (solid line)
in the fully spin-polarized nuclear matter ($x_s=1$).
For the spin-independent parts the parameter-sets of PM1 are used.

\item[Fig. 7]
Density-dependence of the spin-symmetry energies  ($\espsm$) with PM1.
The long-dashed, dashed and solid lines indicate results
for $\CA=0$ (SD1), $=-50$ (SD2) and $=-100$ (SD3), respectively.
The thick and thin lines indicate results at the neutron matter and
at nuclear matter, respectively.

\end{itemize}

\newpage

\begin{table}[t]
\begin{center}
\begin{small}

\begin{tabular}{|c|cccc|}
\hline \hline 
\stret{25pt}
 & \gs & \gv & $B_\sigma$ & $A_\sigma$ \\
\hline \
\stret{20pt}
PM1 & 9.408 & 9.993 & 23.52 & 5.651 \\
\hline
\stret{20pt}
PM4 & 11.05 & 12.64 & 18.89 & 7.158 \\
\hline
\end{tabular}
\caption
{ Parameter sets for the RH calculation in this paper.
In all cases have used \ms = 550 MeV,
\mv = 783 MeV and $C_\sigma$ = 0.}
\end{small}
\end{center}
\end{table}

\vspace*{0.6cm}

\bigskip

\begin{table}[t]
\begin{center}
\begin{small}
\begin{tabular}{|c|cc|cc|}
\hline \hline 
\stret{25pt}
 & \multicolumn{2}{|c|}{PM1} & \multicolumn{2}{|c|}{PM4}  \\
\hline \
\stret{25pt}
 & \CA & \CT  & \CA & \CT  \\
\hline \
\stret{20pt}
SD1 & 0 & 9.993 & 0 & 1.1365 \\
\hline
\stret{20pt}
SD2 & -50 & 56.10 & -50 & 48.32 \\
\hline
\stret{20pt}
SD3 & -100 & 104.5 & -50 & 96.27 \\
\hline
\end{tabular}
\caption
{Parameter sets for the spin-dependent interactions in this paper.}
\end{small}
\end{center}
\end{table}


\begin{thebibliography}{10}

\bibitem{magnetar}
C. Kouveliotou et al., Nature {\bf 393} (1998) 235.\\
K. Hurley et al., Astrophys. J. {\bf 510} (1999) L111. 

\bibitem{SP-QM}
T. Tatsumi, Phys. Lett. {\bf B}, in press; 
hep-ph/9910470(KUNS-1611);\\ 
nucl-th/0002014(KUNS-1636); astro-ph/0004062(KUNS-1656).

\bibitem{blo}
F. Bloch, Z. Phys. {\bf 57} (1929) 545.

\bibitem{Pand}
V.R. Pandharipande, V.K. Garde and J.K. Srivastava,Phys. Lett. {\bf
B38} (1972) 485.

\bibitem{Niem1}
R. Niembro, S. Marcos, M.L. Quelle and J. Navarro, 
Phys. Lett. {\bf B249} (1990) 373.

\bibitem{Niem2}
S. Marcos, R. Niembro and M.L. Quelle
Phys. Lett. {\bf B271} (1991) 277.

\bibitem{RHF} C. J. Horowitz and B. D. Serot, Nucl. Phys. 
{\bf A399} (1983) 529.

\bibitem{Serot} B.D. Serot and J. D. Walecka, The relativistic Nuclear 
Many Body Problem.  In J. W. Negele and E. Vogt, editors, 
$Adv. Nucl. Phys. {\bf Vol. 16}$, page 1, Plenum Press, 1986, 
and reference therein.

\bibitem{soutome} K. Soutome, T. Maruyama, K. Saito,
Nucl. Phys. {\bf 507} (1990) 731. 

\bibitem{TOMO1} T. Maruyama, B. Bl\"attel, W. Cassing, A. Lang, U. Mosel,
K. Weber, Phys. Lett {\bf B297} (1992) 228;\\
T. Maruyama, W. Cassing, U. Mosel, S. Teis and K. Weber, 
Nucl. Phys {\bf A552} (1994) 571.

\bibitem{K-con}
T. Maruyama, H. Fujii,  T. Muto and T. Tatsumi, 
Phys. Lett. {\bf B337} (1994) 19; \\
H. Fujii, T. Maruyama, T. Muto and T. Tatsumi, 
Nucl. Phys. {\bf A597} (1996) 645.

\bibitem{Muether}
H. M\"uther, Prog. Prat. Nucl. Phys. {\bf 14} (1984) 123. 

\bibitem{Chabanat} E.Chabanat et al., Nucl. Phys. {\bf A627} (1997) 710.

\bibitem{TomoTsuka}
T. Maruyama, T.-S. Saito and T. Tsukamoto, Prog. Theo. Phys. 
Vol.{\bf 82} (1989) 1009. 

\bibitem{Navarro} J. Navarro, E.S. Hern\,andez, and D. Vautherin,
hep-ph/901311. 

\bibitem{KLW1} K. Weber, B. Bl\"attel, W. Cassing, H.-C. D\"onges,
V. Koch, A. Lang and U. Mosel, Nucl. Phys. {\bf A539} (1992) 713.

\bibitem{FV} T. Maruyama and S. Chiba,
Phys. Rev. {\bf C61}, 037301-1 (2000).

\bibitem{ushi} S. Ushida and T. Tatsumi, in preparation.


\end{thebibliography}
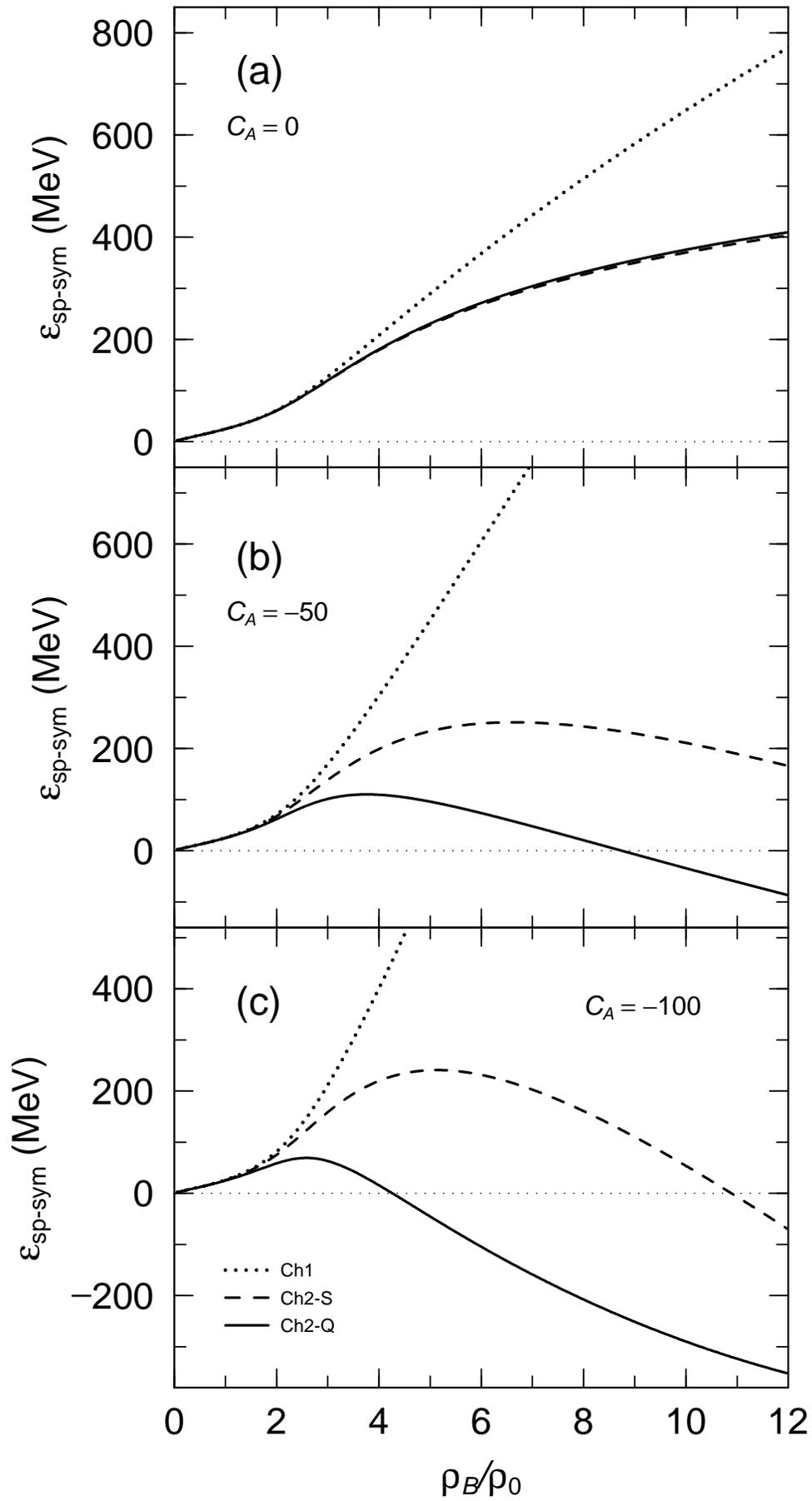
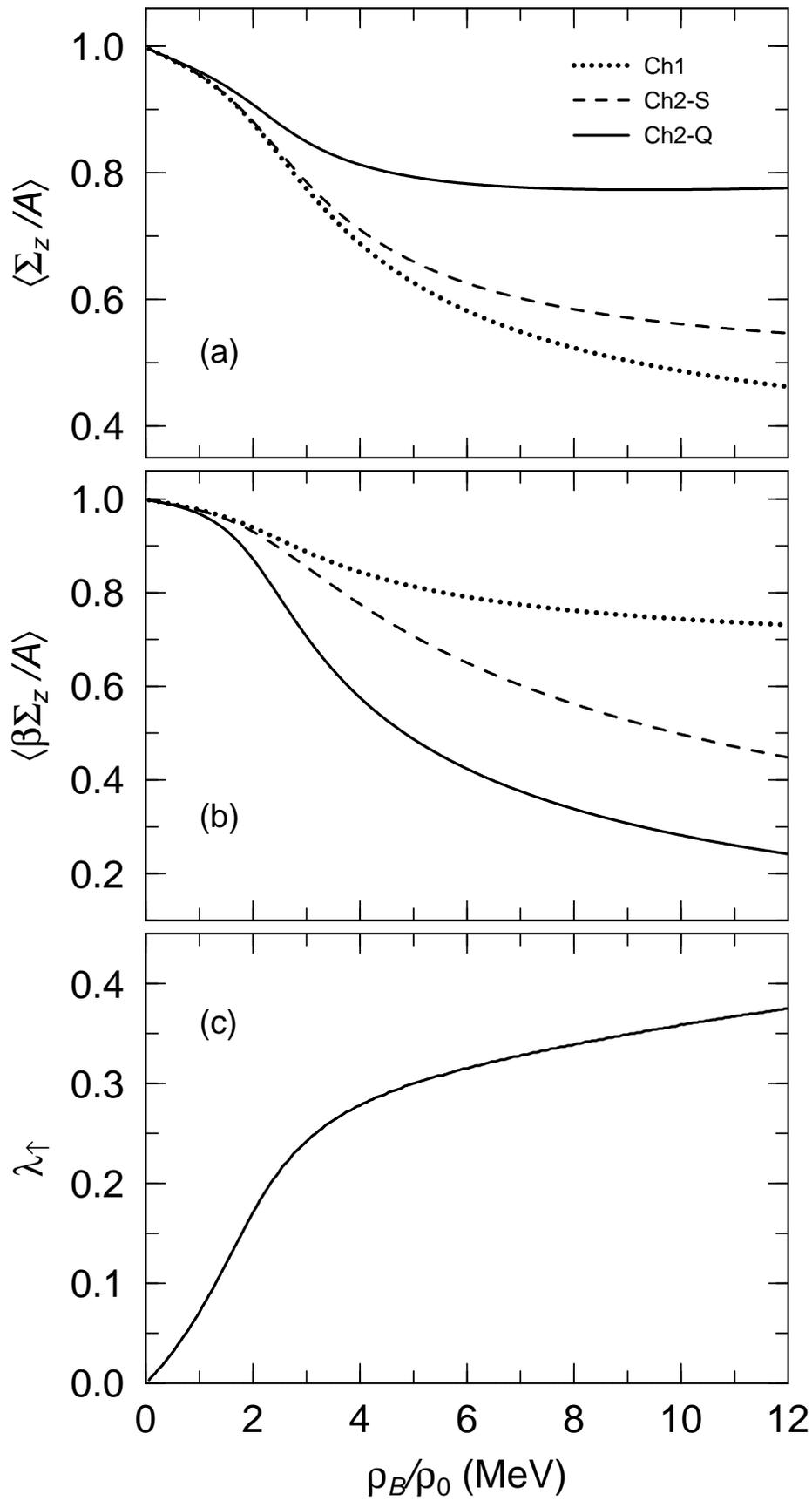
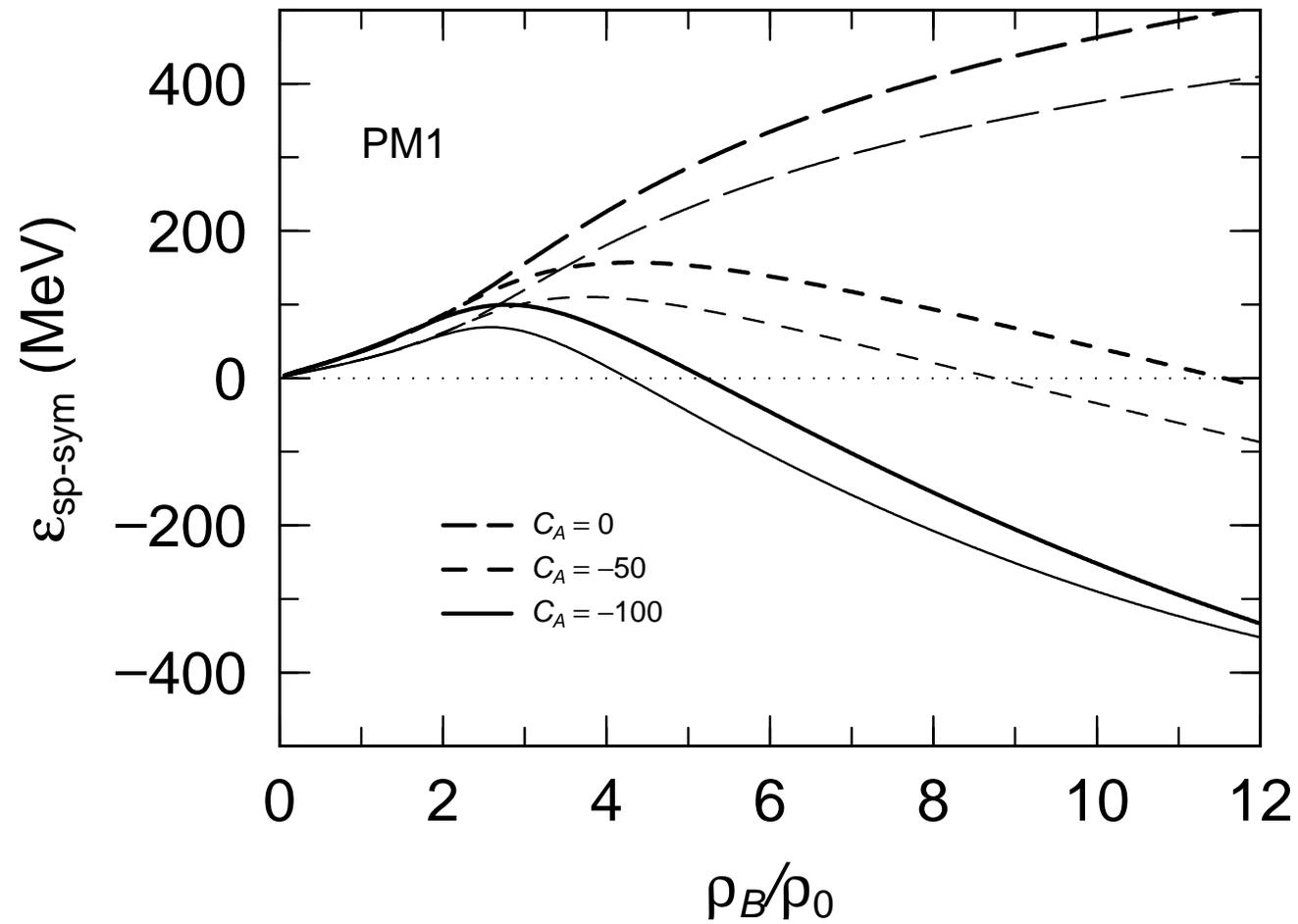
\end{document}